\documentclass[11pt,aps,a4paper,eqsecnum,amsmath,amssymb
              ,nofootinbib,longbibliography
              ,notitlepage
              ]{revtex4-1}

\pdfoutput=1

\usepackage{color}
\usepackage{amsfonts}
\usepackage{graphicx}
\usepackage{amsmath}

\begin{document}

\preprint{}
\title{%
      Condensation of non-Abelian $SU(3)_{N_f}$ anyons in\\ a one-dimensional
      fermion model
}

\author{Daniel Borcherding}
\author{Holger Frahm}
\affiliation{%
  Institut f\"ur Theoretische Physik, Leibniz Universit\"at Hannover,
  Appelstra\ss{}e 2, 30167 Hannover, Germany}

\date{August 17, 2018}

\begin{abstract}
  The color excitations of interacting fermions carrying an $SU(3)$ color and
  $U(N_f)$ flavor index in one spatial dimension are studied in the framework
  of a perturbed $SU(3)_{N_f}$ Wess-Zumino-Novikov-Witten model.  Using Bethe
  ansatz methods the low energy quasi-particles are found to be massive
  solitons forming $SU(3)$ quark and antiquark multiplets.  In addition to the
  color index the solitons carry an internal degree of freedom with
  non-integer quantum dimension.  These zero modes are identified as
  non-Abelian anyons satisfying $SU(3)_{N_f}$ fusion rules.  Controlling the
  soliton density by external fields allows to drive the condensation of these
  anyons into various collective states.  The latter are described by
  parafermionic cosets related to the symmetry of the system.  Based on the
  numerical solution of the thermodynamic Bethe ansatz equations we propose a
  low temperature phase diagram for this model.
\end{abstract}

\maketitle


\section{Introduction}
The remarkable properties of topological states of matter which cannot be
characterized by a local order parameter but rather through their global
entanglement properties have attracted tremendous interest in recent years.
One particular consequence of the non-trivial bulk order is the existence of
fractionalized quasi-particle excitations with unconventional statistics,
so-called non-Abelian anyons.
The ongoing search for physical realizations of these objects is driven by the
possible utilization of their exotic properties, in particular in the quest
for reliable quantum computing where the topological nature of non-Abelian
anyons makes them a potentially promising resource \cite{Kita03,NSSF08}.
Candidate systems supporting excitations with fractionalized zero energy
degrees of freedom are the topologically ordered phases of two-dimensional
quantum matter such as the fractional quantum Hall states or $p+ip$
superconductors \cite{MoRe91,ReRe99,ReGr00}.  In these systems the presence of
gapped non-Abelian anyons in the bulk leads to anomalous physics at the edges
of the probe or at boundaries between phases of different topological order.

To characterize the latter it is essential to understand the properties of an
ensemble of anyons where interactions lift the degeneracies of the anyonic
zero modes and correlated many-anyon states are formed.  One approach towards
a classification of the possible collective states of interacting anyons has
been to study effective lattice models
\cite{FTLT07,GATH13,Finch.etal14,FiFF14,BrFF16,VeJS17,FiFF18}.  Here the local
degrees of freedom are objects in a braided tensor category with operations
describing their fusion and braiding.  Note that both the Hilbert space of the
many-anyon system and the possible local interactions in the lattice model are
determined by the fusion rules.
These models allow for studies of many interacting anyons in one spatial
dimension, both using numerical or, after fine-tuning of couplings to make
them integrable, analytical methods thereby providing important insights into
the collective behaviour of non-Abelian anyons.  However, the question of how
these degrees of freedom can be realized in a microscopic physical system is
beyond its scope of this approach.  In addition, the effective anyon models do
not contain parameters, e.g.\ external fields, which would allow to drive a
controlled transition from a phase with isolated anyons into a condensate.

Here we address these questions starting from a particular one-dimensional
system of fermions: it is well known that strong correlations together with
the quantum fluctuations in such systems may lead to the fractionalization of
the elementary degrees of freedom of the constituents, the best-known example
being spin-charge separation in correlated electron systems such as the
one-dimensional Hubbard model \cite{HUBBARD}.  A similar phenomenon can be
observed in a system of spin-$1/2$ electrons with an additional orbital degree
of freedom which is integrable by Bethe ansatz methods \cite{Tsve14a,BoFr18}:
in the presence of a particularly chosen interaction the elementary
excitations in the spin sector of this system are massive solitons (or kinks)
connecting the different topological ground states of the model.  On these
kinks there are localized zero energy modes which, based on the exact low
temperature thermodynamics, have been identified as spin-$1/2$ anyons
satisfying $SU(2)_k$ fusion rules.  The mass and density of the kinks (and
therefore the anyons) can be controlled by the external magnetic field applied
in the underlying electron system.  This allows to study the condensation of
the anyons into a the collective phase described by a parafermionic conformal
field theory.

Below we extend this work by considering fermions forming an $SU(3)$ 'color'
and an additional $U(N_f)$ flavor multiplet.  We focus on the lowest energy
excitations in the color subsector of the model in the presence of external
fields coupled to the Cartan generators of the global $SU(3)$ symmetry.  The
spectrum of quasi-particles and the low temperature thermodynamics of the
model are studied using Bethe ansatz methods.  Using a combination of
analytical methods and numerical solution of the nonlinear thermodynamic Bethe
ansatz integral equations we identify anyonic zero modes which are localized
on massive solitons in 'quark' and 'antiquark' color multiplets or bound
states thereof.  For sufficiently strong fields the mass gaps of the solitons
close and the anyonic modes overlap.  The resulting interaction lifts the
degeneracy of the zero modes and the anyons condense into a phase with
dispersing collective excitations whose low energy behaviour is described by
parafermionic cosets.  The transitions between the various topological phases
realized by these anyons are signaled by singularities in thermodynamic
quantities at low temperatures.  These are signatures of anyon condensation,
complementing similar studies of two-dimensional topological systems
\cite{LeWe05,BaSl09,HZSV15,IqDS18}.

\section{Integrability study of perturbed $\mathbf{SU(3)_{N_f}}$ WZNW model}
We consider a system described by fermion fields forming a $SU(3)$ 'color'
multiplet and an auxiliary $U(N_f)$ 'flavor' multiplet.  In the presence of
weak interactions preserving the $U(1)$ charge, color, and flavor symmetries
separately, conformal embedding can be used to split the Hamiltonian into a
sum of three commuting parts describing the fractionalized degrees of freedom
in the collective states \cite{FrMSbook96}.  The non-Abelian color degrees of
freedom are described in terms of a critical $SU(3)_{N_f}$
Wess-Zumino-Novikov-Witten model perturbed by current-current interactions
with Hamiltonian density
\begin{equation}
  \label{NfModel}
  \mathcal{H} = \frac{2\pi}{N_f+3} \left(:\mathbf{J}\cdot\mathbf{J}: +
    :\bar{\mathbf{J}}\cdot \bar{\mathbf{J}}:\right) +
  \lambda_\parallel\sum_{i=1}^{2}H^{i}\bar{H}^{i} + \lambda_\perp\sum_{\alpha
    >0}\frac{|\alpha|^2}{2} \left(
    E^{\alpha}\bar{E}^{-\alpha}+E^{-\alpha}\bar{E}^{\alpha} \right)\,. 
\end{equation}
Here $\mathbf{J}$ and $\bar{\mathbf{J}}$ are the right- and left-moving
$SU(3)_{N_f}$ Kac-Moody currents.  In terms of the corresponding fermion
fields their components are
\begin{align*}
  H^{i}&=R^{\dagger}_{a k}(h^{i})_{ab}R_{b k}\,,\qquad 
         E^{\alpha}=R^{\dagger}_{a k}(e^{\alpha})_{ab}R_{b k}\,,\\
  \bar{H}^{i}&=L^{\dagger}_{a k}(h^{i})_{ab}L_{b k}\,,\qquad 
               \bar{E}^{\alpha}=L^{\dagger}_{a k}(e^{\alpha})_{ab}L_{b k}\,,
\end{align*}
where $h^{i}$ $(i=1,2)$ are the generators of the $SU(3)$ Cartan subalgebra
and $e^{\alpha}$ denote the ladder operator for the root $\alpha$ in the
Cartan-Weyl basis ($1\le a,b\le3$ and $1\le k\le N_f$ are color and flavor
indices, respectively).

The spectrum of (\ref{NfModel}) can be obtained using Bethe ansatz methods,
see e.g.\ \cite{FaRe86,BaTs86}, based on the observation that the underlying
structures coincide with those of an integrable deformation of the $SU(3)$
magnet with Dynkin label $(N_f,0)$
\cite{KuRe81,PeSc81,KuRS81,BaVV82,KuRe83,AnJo84b}.  Specifically, placing
$\mathcal{N}=\sum \mathcal{N}_\alpha$ fermions into a box of length $L$ with
periodic boundary conditions and applying magnetic fields $H_1$, $H_2$ coupled
to the conserved charges the energy eigenvalues in the sector with
$\mathcal{N}_1\ge \mathcal{N}_2 \ge \mathcal{N}_3$ are
\begin{equation}
  \label{energy}
  E=\frac{\mathcal{N}}{L}\sum_{\alpha=1}^{N_1}\sum_{\tau=\pm
    1}\frac{\tau}{2\rm{i}}\ln\left(\frac{\sinh(\frac{\pi}{2p_0}(\lambda^{(1)}_\alpha+
      \tau/g-N_f\rm{i}))}{\sinh(\frac{\pi}{2p_0}(\lambda^{(1)}_\alpha+\tau/g+N_f\rm{i}))}
  \right)+N_1 H_1 +N_2 H_2
  -\mathcal{N}\left(\frac{2}{3}H_1+\frac{1}{3}H_2\right)\,, 
\end{equation}
where $N_1=\mathcal{N}_2+\mathcal{N}_3$, $N_2=\mathcal{N}_3$, and the
parameters $g$ and $p_0>N_f$ are functions of the coupling constants
$\lambda_{\parallel}$ and $\lambda_{\perp}$. The relativistic invariance of
the fermion model is broken by the choice of boundary conditions but will be
restored later by considering observables in the scaling limit
$L,\mathcal{N}\to\infty$ and $g\ll1$ such that the mass of the elementary
excitations is small compared to the particle density $\mathcal{N}/L$.

The energy eigenvalues (\ref{energy}) are parameterized by complex parameters
$\lambda^{(m)}_{\alpha}$ with $\, \alpha=1,\dots,N_m$ and $m=1,2$ solving the
hierarchy of Bethe equations (cf.\ Refs.~\cite{PeSc81,Schu83,VeLo91,Lopes92}
for the magnet in the fundamental representation, $N_f=1$)
\begin{equation}
  \label{betheeq}
  \begin{aligned}
    \prod_{\tau=\pm 1}e_{N_f}(\lambda^{(1)}_\alpha+\tau/g)^{\mathcal{N}/2}&=\prod_{\beta \neq \alpha}^{N_1}e_2(\lambda^ {(1)}_\alpha - \lambda^{(1)}_{\beta})\prod_{\beta=1}^{N_2}e_{-1}(\lambda^{(1)}_\alpha-\lambda^{(2)}_\beta)\,,\quad \alpha=1,\ldots, N_1\\
    \prod_{\beta=1}^{N_1}e_{1}(\lambda^{(2)}_\alpha-\lambda^{(1)}_\beta)&=\prod_{\beta
      \neq \alpha}^{N_2}e_2(\lambda^{(2)}_\alpha-\lambda^{(2)}_\beta)\,,\quad
    \alpha=1,\ldots,N_2
  \end{aligned}
\end{equation}
where
$e_k(x)=\sinh\left(\frac{\pi}{2p_0}(x+{\rm i}k)\right)/
\sinh\left(\frac{\pi}{2p_0}(x-{\rm i}k)\right)$. Based on these equations the
thermodynamics of the model can be studied provided that the solutions to the
Bethe equations describing the eigenstates in the limit $\mathcal{N}\to\infty$
are known.  Here we argue that the root configurations corresponding to the
ground state and excitations relevant for the low temperature behaviour of
(\ref{NfModel}) can be built based on the string hypothesis for the
$U_q(sl(2))$ models \cite{TaSu72}, i.e.\ that in the thermodynamic limit the
Bethe roots $\lambda_\alpha^{(m)}$ on both levels $m=1,2$ can be grouped into
'$j$-strings' of length $n_j$ and with parity $v_{n_j}\in\{\pm 1\}$
\begin{equation}
  \label{string}
  \lambda^{(m)}_{j,\alpha,k}=\lambda^{(m)}_{j,\alpha} + i\left(n_j+1-2k+\frac{p_0}{2}(1-v_{N_f}v_{n_j})\right)\,, \quad \quad k=1,\dots,n_j,\quad m=1,2\,,
\end{equation}
with real centers $\lambda^{(m)}_{j,\alpha} \in \mathbb{R}$.  The allowed
lengths and parities depend on the parameter $p_0$.  In the following we
further simplify our discussion by assuming $p_0=N_f+1/\nu$ with integer
$\nu>2$ where the following $N_f+\nu$ different string configurations
contribute to the low temperature thermodynamics:
\begin{itemize}
\item $j_0 = N_f+\nu$ with $(n_{j_0},v_{n_{j_0}}) = (N_f,+1)$,
\item $j_1\in \{N_f,N_f+1,\ldots N_f+\nu-1\}$ with
  $(n_{j_1},v_{n_{j_1}})=((j_1-N_f)N_f+1,(-1)^{j_1+N_f+1})$,
\item $j_2\in \{1,2,\ldots,N_f-1\}$ with $(n_{j_2},v_{n_{j_2}}) = (j_2,+1)$.
\end{itemize}
Within the root density approach the Bethe equations are rewritten as coupled
integral equations for the densities of these strings \cite{YaYa66b}.  For
vanishing magnetic fields one finds that the Bethe root configuration
corresponding to the lowest energy state is described by finite densities of
$j_0$-strings on both levels $m=1,2$.
The elementary excitations above this ground state are of three types: similar
as in the isotropic magnet \cite{Joha86} there are solitons or 'quarks' and
'antiquarks' corresponding to holes in the distributions of $j_0$-strings on
level $m=1,2$.  They carry quantum numbers in the fundamental representations
$(1,0)$ and $(0,1)$ of $SU(3)$, respectively (independent of the
representation $(N_f,0)$ used for the construction of the spin chain).  The
$\nu$ different types of $j_1$-strings are called 'breathers'.  Finally, there
are auxiliary modes given by $j_2$-strings.  The densities
$\rho^{(m)}_{j}(\lambda)$ of these excitations (and $\rho^{h(m)}_{j}(\lambda)$
for the corresponding vacancies) satisfy the integral equations ($a\ast b$
denotes the convolution of $a$ and $b$)
\begin{equation}
  \label{densities2}
  \rho^{h(m)}_k(\lambda)={\rho}^{(m)}_{0,k}(\lambda)-\sum_{\ell=1}^{2}\sum_j \left(B^{(m,\ell)}_{kj}\ast \rho^{(\ell)}_j\right)(\lambda)\,, \qquad m=1,2\,,
\end{equation}
see Appendix~\ref{app:onTBA}.
As mentioned above relativistic invariance is restored in the scaling limit
$g\ll 1$ where the solitons are massive particles with bare densities
${\rho}^{(m)}_{0,j_0} (\lambda)$ and bare energies $\epsilon^{(m)}_{0,j_0}$
\begin{equation}
  \label{bare_kink}
  \begin{aligned}
    {\rho}^{(m)}_{0,j_0}(\lambda)&\stackrel{g\ll 1}{\simeq} \frac{M_0}{6}\cosh(\pi \lambda/3)\,,\\
    {\epsilon}^{(m)}_{0,j_0}(\lambda)&\stackrel{g\ll 1}{\simeq} M_0\cosh(\pi
    \lambda/3)-
    \begin{cases}
      (Z_{1}H_1+Z_2H_2)\,,\quad m=1,\\
      (Z_{2}H_1+Z_1H_2)\,,\quad m=2\,.
    \end{cases}
  \end{aligned}
\end{equation}
Here $M_{0}\equiv M_{j_0}= (2\mathcal{N}/L) \sin(\pi/3) e^{-\frac{\pi}{3g}}$
is the soliton mass while $Z_1=\frac{2}{3}(1+N_f\nu)$ and
$Z_2=\frac{1}{3}(1+N_f\nu)$ parameterize the $SU(3)$ charges of these
excitations corresponding to the highest weight states in the quark and
antiquark representation.  Similarly, breathers have bare densities and
energies
\begin{equation}
  \label{bare_breather}
  \begin{aligned}
    {\rho}^{(m)}_{0,j_1}(\lambda)&\stackrel{g\ll 1}{\simeq} \frac{M_{j_1}}{6}\cosh(\pi \lambda/3)\,,\\
    {\epsilon}^{(m)}_{0,j_1}(\lambda)&\stackrel{g\ll 1}{\simeq}
    M_{j_1}\cosh(\pi \lambda /3)-
    \begin{cases}
      (-1)^{m}Z_2(H_1-H_2)\,, &\text{if } j_1=N_f+\nu-1,\\
      (Z_2H_1+Z_1H_2)\,, &\text{if } 0\le j_1-N_f < \nu-1\,,\, m=1,\\
      (Z_1H_1+Z_2H_2)\,, &\text{if } 0\le j_1-N_f <\nu-1\,,\, m=2\,,
    \end{cases}
  \end{aligned}
\end{equation}
with masses 
\begin{equation}
  \begin{aligned}
    M_{j_1} &\equiv 
    \begin{cases}
      2M_{0}\,\sin\left((j_1-N_f+1)\frac{\pi}{3\nu}+\frac{\pi}{6}\right)\,, & \text{if }0\le j_1 -N_f < \nu-1\,,\\
      M_{0}\,, & \text{if }j_1= N_f+\nu-1\equiv \tilde{j}_0\,.
    \end{cases}
  \end{aligned}
\end{equation}
Note that the mass of the breathers with $j_1=\tilde{j}_0$ coincides with that
of the solitons.  The magnetic fields, however, couple to these modes in a
different way, indicating that they are descendents of the $SU(3)$ highest
weight states in the quark and antiquark multiplet.  Therefore excitations of
types $j_0$ and $\tilde{j}_0$ will both be labelled solitons (or quarks and
antiquarks for solitons on level $m=1$ and $2$, respectively) below.
The masses and $SU(3)$ charges of the auxiliary modes vanish, i.e.\
${\rho}^{(m)}_{0,j_2}(\lambda) = 0 = {\epsilon}^{(m)}_{0,j_2}(\lambda)$.

The energy density of a macro-state with densities given by (\ref{densities2})
is
\begin{equation}
  \label{energy3}
  \Delta\mathcal{E}=\sum_{m=1}^{2}\sum_{j}\int_{-\infty}^{\infty}\text{d}\lambda\,{\epsilon}^{(m)}_{0,j}(\lambda)\rho^{(m)}_{j}(\lambda)\,.
\end{equation}

\section{Low Temperature Thermodynamics}
Additional insights into the physical properties of the different
quasi-particles appearing in the Bethe ansatz solution of the model
(\ref{NfModel}) can be obtained from its low temperature thermodynamics.  The
equilibrium state at finite temperature is obtained by minimizing the free
energy, $F/\mathcal{N}=\mathcal{E}-T\mathcal{S}$, with the combinatorial
entropy density \cite{YaYa69}
\begin{equation}
  \label{entropy}
  \mathcal{S}=\sum_{m=1}^2 \sum_{j} \int\text{d}\lambda\,
  \left[(\rho^{(m)}_j+\rho^{h(m)}_j)\ln(\rho^{(m)}_j+\rho^{h(m)}_j)-\rho^{(m)}_j\ln\rho^{(m)}_j-\rho^h_j\ln\rho^{h(m)}_j\right]\,. 
\end{equation}

The resulting thermodynamic Bethe ansatz (TBA) equations for the dressed
energies
$\epsilon_j^{(m)}(\lambda) = T\ln\left( \rho_j^{h(m)}(\lambda)/
  \rho_j^{(m)}(\lambda) \right)$ read
\begin{align}
  \label{dressedeinteq}
  &T\ln(1+e^{\epsilon^{(m)}_k/T})={\epsilon}^{(m)}_{0,k}(\lambda) +\sum_{\ell=1}^{2}\sum_{j} B^{(\ell,m)}_{jk}\ast T\ln(1+e^{-\epsilon^{(\ell)}_j/T}).
\end{align}
It is convenient to rewrite the equations for the auxiliary modes
$j\in\{j_2\}$
\begin{equation}
  \label{auxinteq}
  \begin{aligned}
    \epsilon^{(m)}_{j_2}(\lambda)=&s\ast T\ln(1+e^{\epsilon^{(m)}_{j_2+1}/T} )(1+e^{\epsilon^{(m)}_{j_2-1}/T})-s\ast T\ln (1+e^{-\epsilon^{(3-m)}_{j_2}/T})\,,\\ &\qquad j_2=1,\ldots,N_f-2\,,\\
    \epsilon^{(m)}_{N_f-1}(\lambda)=& s\ast T\ln(1+e^{\epsilon^{(m)}_{N_f-2}/T})-s\ast T\ln (1+e^{-\epsilon^{(3-m)}_{N_f-1}/T} )\\
    &+\sum_{j \notin \{j_2\}}C^{(m)}_j\ast T\ln(1+e^{-\epsilon^{(m)}_j/T})\,,
  \end{aligned}
\end{equation}
where $\epsilon^{(m)}_0=-\infty$, $s(\lambda)=\frac{1}{4\cosh{\pi\lambda/2}}$
and the kernels $C^{(m)}_j$ are defined in Appendix~\ref{app:onTBA}. Notice
that the integral equations for the auxiliary modes (\ref{auxinteq}) coincide
with the integral equations of RSOS models of $A_2$ type up to the driving
terms \cite{BaRe90}.
The free energy per particle in terms of the solutions to the TBA equations for the solitons and breathers as
\begin{equation}
  \label{freeenergy}
  \begin{aligned}
    \frac{F}{\mathcal{N}}&\,\,=\,\,-T\sum_{m=1}^{2}\sum_{j \notin\{j_2\}} \int_{-\infty}^{\infty}\text{d}\lambda\, {\rho}^{(m)}_{0,j}(\lambda)\ln(1+e^{-\epsilon^{(m)}_{j}/T})\,\\
    &\stackrel{g\ll 1}{=}-\frac{T}{6}\sum_{m=1}^{2}\sum_{j \notin\{j_2\}} M_{j}\int_{-\infty}^{\infty}\text{d}\lambda\, \cosh(\pi\lambda/3)\ln(1+e^{-\epsilon^{(m)}_{j}/T})\,.
  \end{aligned}
\end{equation}

\begin{figure}
  \includegraphics[width=0.6\textwidth]{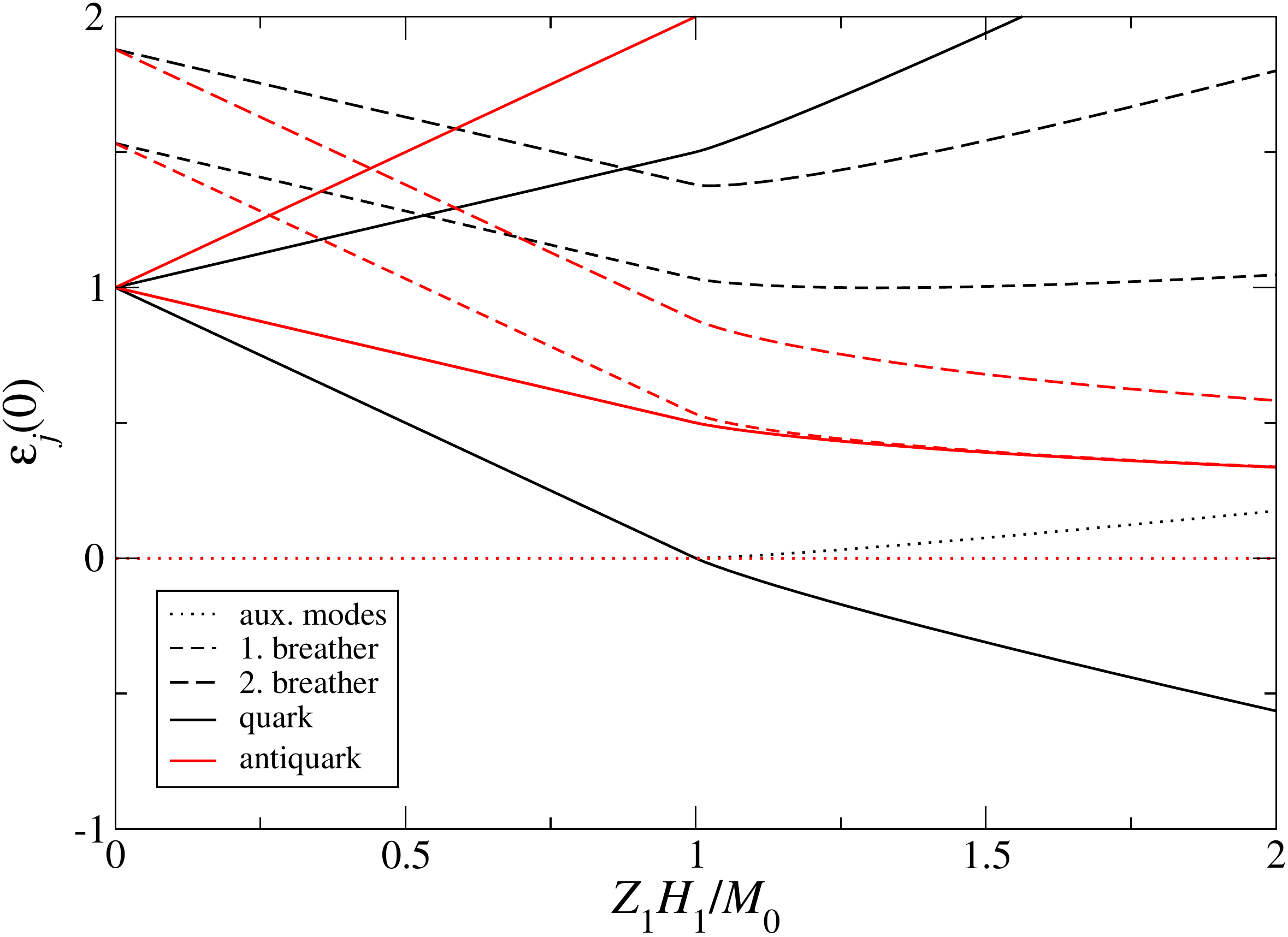}
  \caption{
    The energy gap of the elementary excitations (and Fermi energy of quarks
    in the condensed phase, respectively) $\epsilon^{(m)}_j(0)$ obtained from
    the numerical solution of (\ref{dressedeinteq}) as a function of the field
    $H_1$ for $p_0=2+1/3$ at zero temperature and field $H_2=0$ (gaps on level
    $m=1$ ($2$) are displayed in black (red)). Note that in this case the high
    energy quark and the low energy antiquark levels are twofold degenerate.
    For $Z_1H_1 = M_0$ the quark gap ($\epsilon^{(1)}_{j_0}(0)$) closes and
    the system forms a collective state of these objects.  In this phase the
    degeneracy of the auxiliary modes is lifted.  Increasing the field to
    $Z_1H_1 \gg M_0$ the gaps of the antiquarks ($\epsilon^{(2)}_{j_0}(0)$ and
    $\epsilon^{(2)}_{\tilde{j}_0}(0)$) close. For small fields the low lying
    auxiliary modes are clearly separated from the spectrum of solitons and
    breathers. 
    \label{fig:spec1}}
\end{figure}

\begin{figure}
    \includegraphics[width=0.6\textwidth]{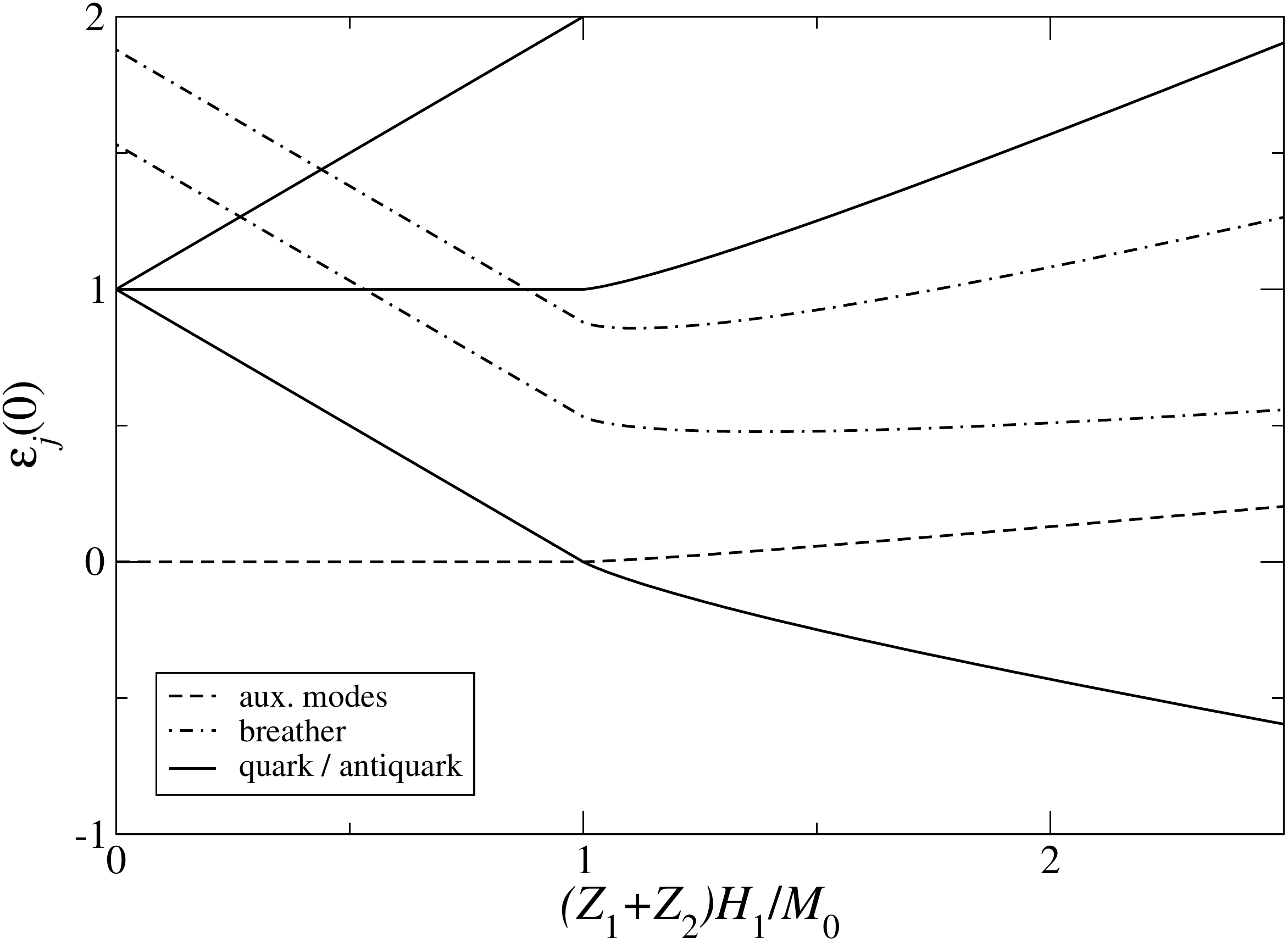}
  
  \caption{
    Same as Fig.~\ref{fig:spec1} but for magnetic fields $H_1\equiv H_2$. In
    this case the elementary excitations for $m=1$ and $2$ are degenerate. 
    \label{fig:spec2}}
\end{figure}

Solving the TBA equations (\ref{dressedeinteq}) we obtain the spectrum of the
model (\ref{NfModel}) for given temperature $T$ and fields.  In the following
we restrict ourselves to the regime $H_1 \geq H_2$ -- exchanging
$H_1\leftrightarrow H_2$ corresponds to interchanging the two levels of the
Bethe ansatz.  From the expressions (\ref{bare_kink}) and
(\ref{bare_breather}) for the bare energies of the elementary excitations we
can deduce the qualitative behaviour of these modes at low temperatures: 
As long as $Z_1H_1+Z_2H_2\lesssim M_{0}$ solitons and breathers remain gapped.
Increasing the fields with $H_1>H_2$ the gap of the quarks ($m=1$ in
(\ref{bare_kink})) closes once $Z_1H_1+Z_2H_2\simeq M_0$.  For larger fields
they condense into a phase with finite density $\rho^{(1)}_{j_0}$ and the
degeneracy of the corresponding zero energy auxiliary modes is lifted.  At
even larger fields the gap of the $SU(3)$ highest weight state of the
antiquark will close, too, and the systems enters a collective phase with a
finite density of quarks and antiquarks.  In Figures~\ref{fig:spec1} and
\ref{fig:spec2} the zero temperature mass spectrum for the model with $N_f=2$,
$\nu=3$ is shown as function of $H_1$ for $H_2=0$ and $H_1=H_2$,
respectively.\footnote{%
  We note that the highest energy soliton levels are not captured by the
  string hypothesis (\ref{string}).  However, since the gaps of these modes
  grow with the magnetic field they do not contribute to the low temperature
  thermodynamics studied in this paper.}
Note that in the latter case the spectra of elementary excitations on level
$1$ and $2$ coincide, $\epsilon^{(1)}_j\equiv \epsilon^{(2)}_j$ for all $j$.

Based on this picture we now discuss the behaviour of the free energy as
function of the fields at temperatures small compared to the relevant energy
scales, i.e.\ the masses or Fermi energies of the solitons,
$0<T\ll |Z_1H_1+Z_2H_2-M_0|$.

\subsection{Non-interacting solitons}
For fields $Z_1H_1 + Z_2 H_2\lesssim M_0$ solitons and breathers are gapped.
At temperatures small compared to the gaps of the solitons the nonlinear
integral equations (\ref{dressedeinteq}) can be solved by iteration: the
energies of the massive excitations are well described by their first order
approximation \cite{Tsve14a} while those of the auxiliary modes are given by
the asymptotic solution of (\ref{auxinteq}) for $|\lambda|\rightarrow \infty$
\cite{Kiri89}
\begin{equation}
  1+e^{\epsilon^{(m)}_{j_2}/T}=\frac{\sin\left(\frac{(j_2+1)\pi}{N_f+3}\right)\sin\left(\frac{(j_2+2)\pi}{N_f+3}\right)}{\sin\left(\frac{\pi}{N_f+3}\right)\sin\left(\frac{2\pi}{N_f+3}\right)}\,.
\end{equation}
For solitons and breathers this implies
($Q\equiv1+e^{\epsilon^{(1)}_{1}}=\frac{\sin(3\pi
  /(N_f+3))}{\sin(\pi/(N_f+3))}$)
\begin{equation}
  \begin{aligned}
    \epsilon^{(m)}_{j}(\lambda)=\begin{cases}\tilde{\epsilon}^{(0,m)}_j(\lambda)-T\ln Q\qquad &j=j_0,\tilde{j}_0\\
      \tilde{\epsilon}^{(0,m)}_j(\lambda)-T\ln Q^2\quad &j\in\{j_1\}\setminus \{\tilde{j}_0\}
    \end{cases}
  \end{aligned}
\end{equation}
resulting in the free energy
\begin{equation}
  \label{eq:Fideal}
  \begin{aligned}
    \frac{F}{\mathcal{N}}=&-\sum_{m=1}^{2}\sum_{j=j_0,\tilde{j}_0}TQ\int\frac{\text{d}p}{2\pi}\,
    e^{-\left({\epsilon}^{(m)}_{0,j}(0)-p^2/2M_{0}\right)/T}-\sum_{m=1}^{2}\sum_{j\in\{j_1\}\setminus
      \{\tilde{j}_0\}}TQ^2\int\frac{\text{d}p}{2\pi}e^{-\left({\epsilon}^{(m)}_{0,j}(0)-p^2/2M_{j}\right)T}\,. 
  \end{aligned}
\end{equation}
As observed in Refs.~\cite{Tsve14a,BoFr18} each of the terms appearing in this
expression is the free energy of an ideal gas of particles with the
corresponding mass carrying an internal degree of freedom with non-integer
'quantum dimension' $Q$ for the solitons and $Q^2$ for the breathers
($j_1\ne\tilde{j}_0$). Their densities
\begin{equation}
  \begin{aligned}
    &n_{j}=\begin{cases}Q\sqrt{\frac{M_{0}T}{2\pi}}e^{-{\epsilon}^{(m)}_{0,j}(0)/T} \qquad &j=j_0,\tilde{j}_0,\\
      Q^2\sqrt{\frac{M_{j}T}{2\pi}}e^{-{\epsilon}^{(m)}_{0,j}(0)/T} \qquad &j\in \{j_1\}\setminus \{\tilde{j}_0\}
    \end{cases}
  \end{aligned}
\end{equation}
can be controlled by variation of the temperature and the fields, which act as
chemical potentials.

For the interpretation of this observation we consider fields $H_1 >H_2$ and
$Z_1H_1+Z_2H_2\lesssim M_0$ where the dominant contribution to the free energy
is that of the lowest energy quarks, $j=j_0$, $m=1$. Their degeneracy $Q$
coincides with the quantum dimension of the anyons satisfying $SU(3)_{N_f}$
fusion rules with topological charge $[1,0]$ or $[1,1]$ 
\begin{equation}
    d_{N_f}([x_1,x_2]) =
    \frac{\sin\left(\frac{\pi(x_1+2)}{N_f+3}\right)\,
      \sin\left(\frac{\pi(x_2+1)}{N_f+3}\right)\,
      \sin\left(\frac{\pi(x_1-x_2+1)}{N_f+3}\right)}  
    {\sin\left(\frac{\pi}{N_f+3}\right)\, \sin\left(\frac{\pi}{N_f+3}\right)\,
      \sin\left(\frac{2\pi}{N_f+3}\right)}\,, 
\end{equation}
see Refs.~\cite{BaRe90,FrKa15,SchWe16}.  Here $[x_1,x_{2}]$ denotes the Young
diagram with $x_i$ nodes in the $i$-th row.  Following the discussion in
Ref.~\cite{Tsve14a} we interpret this as a signature for the presence of
$SU(3)_{N_f}$ anyonic zero modes bound to the quarks.  The degeneracy $Q^2$ of
the breather can be understood as a consequence of the breather being a bound
state of two quarks, each contributing a factor $Q$ to the quantum dimension:
from the fusion rule for $SU(3)_{N_f}$ $[1,0]$ and $[1,1]$ anyons,
\begin{equation*}
[1,0]\times [1,0]=[1,1]+[2,0]\,,\quad [1,1]\times [1,1]=[1,0]+[2,2]\,,
\end{equation*}
for $N_f>1$, the degeneracy of this bound state is obtained to be $Q^2 = d_{N_f}([1,1])+d_{N_f}([2,0]) = d_{N_f}([1,0])+d_{N_f}([2,2])$.

\subsection{Condensate of quarks}
\label{sectionB}
For fields $Z_1 H_1 \gtrsim M_0-Z_2 H_2 > \frac23 M_0$ the quarks in the
$SU(3)$ highest weight state form a condensate, while the contribution to the
free energy of the other solitons and the breathers can be neglected.  For
large fields $Z_1 H_1 \gg M_0 - Z_2 H_2$ the low temperature thermodynamics in
this regime can be studied analytically: following
\cite{KiRe87b} we observe that the dressed energies and
densities can be related as
\begin{equation}
  \label{derelations}
  \begin{aligned}
    \rho^{(m)}_j(\lambda)&=(-1)^{\delta_{j\in\{j_2\}}}\frac{1}{2\pi}\frac{\text{d}\epsilon^{(m)}_j(\lambda)}{\text{d}\lambda} f\left(\frac{\epsilon^{(m)}_j(\lambda)}{T}\right),\\
    \rho^{h(m)}_j(\lambda)&=(-1)^{\delta_{j\in\{j_2\}}}\frac{1}{2\pi}\frac{\text{d}\epsilon^{(m)}_j(\lambda)}{\text{d}\lambda} \left(1-f\left(\frac{\epsilon^{(m)}_j(\lambda)}{T}\right)\right)\,,
  \end{aligned}
\end{equation}
for $\lambda> \lambda_\delta$ with a sufficiently large
$\exp(\pi\lambda_\delta/2)\gg1$.  $f(\epsilon)= 1/(1+e^\epsilon)$ is the Fermi
function.  Inserting this into (\ref{entropy}) we get
($\phi^{(m)}_j=\epsilon^{(m)}_j/T$)
\begin{equation}
  \begin{aligned}
    \label{entropy2}
    \mathcal{S}=&-\frac{T}{\pi}\sum_{m,j}(-1)^{\delta_{j\in\{j_2\}}} \int_{\phi^{(m)}_j(\lambda_\delta)}^{\phi^{(m)}_j(\infty)}\text{d}\phi^{(m)}_j\, \left[f(\phi^{(m)}_j)\ln f(\phi^{(m)}_j)+(1-f(\phi^{(m)}_j))\ln(1-f(\phi^{(m)}_j))\right]\,\\
    &+\sum_{m,j}\mathcal{S}^{(m)}_j(\lambda_\delta)\, ,\\
    \mathcal{S}^{(m)}_j&(\lambda_\delta)\equiv \int_{-\lambda_\delta}^{\lambda_\delta}\text{d}\lambda\, \left[(\rho^{(m)}_j+\rho^{h(m)}_j)\ln(\rho^{(m)}_j+\rho^{h(m)}_j)-\rho^{(m)}_j\ln\rho^{(m)}_j-\rho^{h(m)}_j\ln\rho^{h(m)}_j\right]\,.
  \end{aligned}
\end{equation}
The integrals over $\phi^{(m)}_j$ can be performed giving
\begin{align}
  \label{entropyassymp}
  &\mathcal{S}=\sum_{m,j}\mathcal{S}^{(m)}_j(\lambda_\delta)-\frac{2T}{\pi}\sum_{m,j}(-1)^{\delta_{j\in\{j_2\}}}[L(f(\phi^{(m)}_j(\infty))-L(f(\phi^{(m)}_j(\lambda_\delta)))].
\end{align}
in terms of  the Rogers dilogarithm $L(x)$
\begin{equation}
  L(x)=-\frac{1}{2}\int_{0}^{x}\text{d}y\, \left(\frac{\ln y}{1-y}+\frac{\ln (1-y)}{y}\right)\,.
\end{equation}
In the regime considered here, i.e.\ $T\ll Z_1H_1+Z_2H_2-M_0$ and
$\log((Z_1H_1+Z_2H_2)/M_0)>\lambda_\delta\gg1$, we conclude from
(\ref{dressedeinteq}),(\ref{auxinteq}) that
\begin{equation}
  \label{ccondition}
  \begin{aligned}
    &f(\phi^{(m)}_{j_0}(\lambda_\delta))= \begin{cases}
      1 & m=1\\
      0 & m=2
    \end{cases},
    & f(\phi^{(m)}_{j_0}(\infty))&=0,\\
    &f(\phi^{(m)}_{j_1}(\lambda_\delta))=0, &  f(\phi^{(m)}_{j_1}(\infty))&=0,\\
    &f(\phi^{(m)}_{j_2}(\lambda_\delta))=
    \begin{cases}
      0 & m=1\\
      \left(\frac{\sin(\frac{\pi}{N_f+2})}{\sin(\frac{\pi(j_2+1)}{N_f+2})}\right)^2 & m=2
    \end{cases},&
    f(\phi^{(m)}_{j_2}(\infty))&=\frac{\sin(\frac{\pi}{N_f+3})\sin(\frac{2\pi}{N_f+3})}{\sin(\frac{j_2\pi}{N_f+3})\sin(\frac{(j_2+1)\pi}{N_f+3})},
  \end{aligned}
\end{equation}
and therefore
\begin{equation}
  \rho^{h(1)}_{j_0}(\lambda)=\rho^{(2)}_{j_0}(\lambda)=\rho^{(m)}_{j_1}(\lambda)=\rho^{(1)}_{j_2}=0 \quad 
\end{equation}
for $|\lambda|<\lambda_\delta$.  Using
$\rho^{(2)}_{j_2}(\lambda)=e^{-\epsilon^{(2)}_{j_2}/T}\rho^{h(2)}_{j_2}(\lambda)$
the integral equations (\ref{densities2}) for $\rho^{h(2)}_{j_2}$ simplify in
this regime to
\begin{equation}
  \rho^{h(2)}_{j_2}=-\sum_{k_2}B^{(2,2)}_{j_2k_2}\ast e^{-\epsilon^{(2)}_{k_2}/T}\rho^{h(2)}_{k_2}\qquad \text{for }|\lambda|<\lambda_\delta,
\end{equation}
we conclude that $\rho^{h(2)}_{j_2}\rightarrow 0,$ $\rho^{(2)}_{j_2}\rightarrow 0$ such that $\rho^{(2)}_{j_2}/\rho^{h(2)}_{j_2}=e^{-\epsilon^{(2)}_{j_2}/T}=\text{const}$ for $|\lambda|<\lambda_\delta$ Consequently, we get $\mathcal{S}^{(m)}_j(\lambda_\delta)=0$ for all $j,m$. Using $L(1)=\pi^2/6$ and the relation with ($N,N_f\geq 2$)
\begin{equation}
  \label{RogerF}
  \sum_{m=1}^{N-1}\sum_{j_2=1}^{N_f-1}L\left(\frac{\sin(\frac{(N-m)\pi}{N_f+N})\sin(\frac{m\pi}{N_f+N})}{\sin(\frac{(N+j_2-m)\pi}{N_f+N})\sin(\frac{(j_2+m)\pi}{N_f+N})} 
  \right)=\frac{\pi^2}{6}\left(\frac{N_f(N^2-1)}{N_f+N}-(N-1) \right) 
\end{equation}
for $N=2,3$ we get the following low-temperature behavior of the entropy
\begin{equation}
  \label{Entropy_CFT1}
  \mathcal{S}=\frac{\pi}{3}\left(\frac{8N_f}{N_f+3}-\frac{3N_f}{N_f+2}\right)T.
\end{equation}
This is consistent with an effective description of the low energy collective
modes in this regime through the coset $SU(3)_{N_f}/SU(2)_{N_f}$ conformal
field theory with central charge
\begin{equation}
  c=\frac{8N_f}{N_f+3}-\frac{3N_f}{N_f+2}.
\end{equation}
Using the conformal embedding \cite{CFKM00} (see also \cite{HuNY90})
\begin{equation}
  \label{embedding1}
  \frac{SU(3)_{N_f}}{SU(2)_{N_f}}=U(1)+\frac{Z_{SU(3)_{N_f}}}{Z_{SU(2)_{N_f}}},
\end{equation}
where $Z_{SU(N)_{N_f}}=SU(N)_{N_f}/U(1)^N$ denotes generalized parafermions
\cite{Gepner87}, the collective modes can equivalently be described by a
product of a free $U(1)$ boson contributing $c=1$ to the central charge and a
parafermion coset $Z_{SU(3)_{N_f}}/Z_{SU(2)_{N_f}}$ contributing 
\begin{equation}
  c=\frac{8N_f}{N_f+3}-\frac{3N_f}{N_f+2}-1=\frac{6(N_f-1)}{N_f+3}-\frac{2(N_f-1)}{N_f+2}\,.
\end{equation}

To study the transition from the gas of free anyons to the condensate of
quarks described by the CFT (\ref{embedding1}) at intermediate fields
$Z_1 H_1 \gtrsim M_0-Z_2 H_2$ we have solved the TBA equations
(\ref{dressedeinteq}) numerically.  Similar as in Ref.~\cite{BoFr18} this can
be done choosing suitable initial distributions and iterating the integral
equations for given fields $H_1,H_2$, temperature $T$ and anisotropy parameter
$p_0$.

Using (\ref{freeenergy}) the entropy can be computed from the numerical data
as
\begin{equation}
    \label{entropy_num}
    \begin{aligned} \mathcal{S} &=
-\frac{\text{d}}{\text{d}T}\frac{F}{\mathcal{N}}\\ &= \sum_{m,j\notin\{j_2\}}
\frac{M_j}{4}\int\text{d}\lambda\, \cosh\left(\frac{\pi\lambda}{2}\right)
\left(\log\left(1+e^{-\epsilon^{(m)}_j/T}\right)+\left(\frac{\epsilon^{(m)}_j}{T}-\frac{\text{d}}{\text{d}T}\epsilon^{(m)}_j\right)\left(1+e^{\epsilon^{(m)}_j/T}\right)^{-1}
\right).
    \end{aligned}
\end{equation}
From the numerical solution of the TBA equations one finds that the low energy
behaviour is determined by the quarks and the auxiliary modes on the first
level which propagate with Fermi velocities
    \begin{equation} v^{(1)}_{\text{quark}} = \left.\frac{\partial_\lambda
\epsilon^{(1)}_{j_0}} {2\pi\rho^{h(1)}_{j_0}}\right|_{\Lambda} \,,\qquad
v^{(1)}_{pf} = -\left.\frac{ \partial_\lambda \epsilon^{(1)}_{j_2}(\lambda)
}{2\pi\rho^{h(1)}_{j_2}(\lambda)}\right|_{\lambda\to\infty}\,,
\end{equation} where $\Lambda$ is defined by
$\epsilon^{(1)}_{j_0}(\Lambda)=0$.  Note that $v^{(1)}_{pf}$ is the same for
all auxiliary modes $(j\in\{j_2\}, m=1)$

\begin{figure}[ht]
  \includegraphics[width=0.6\textwidth]{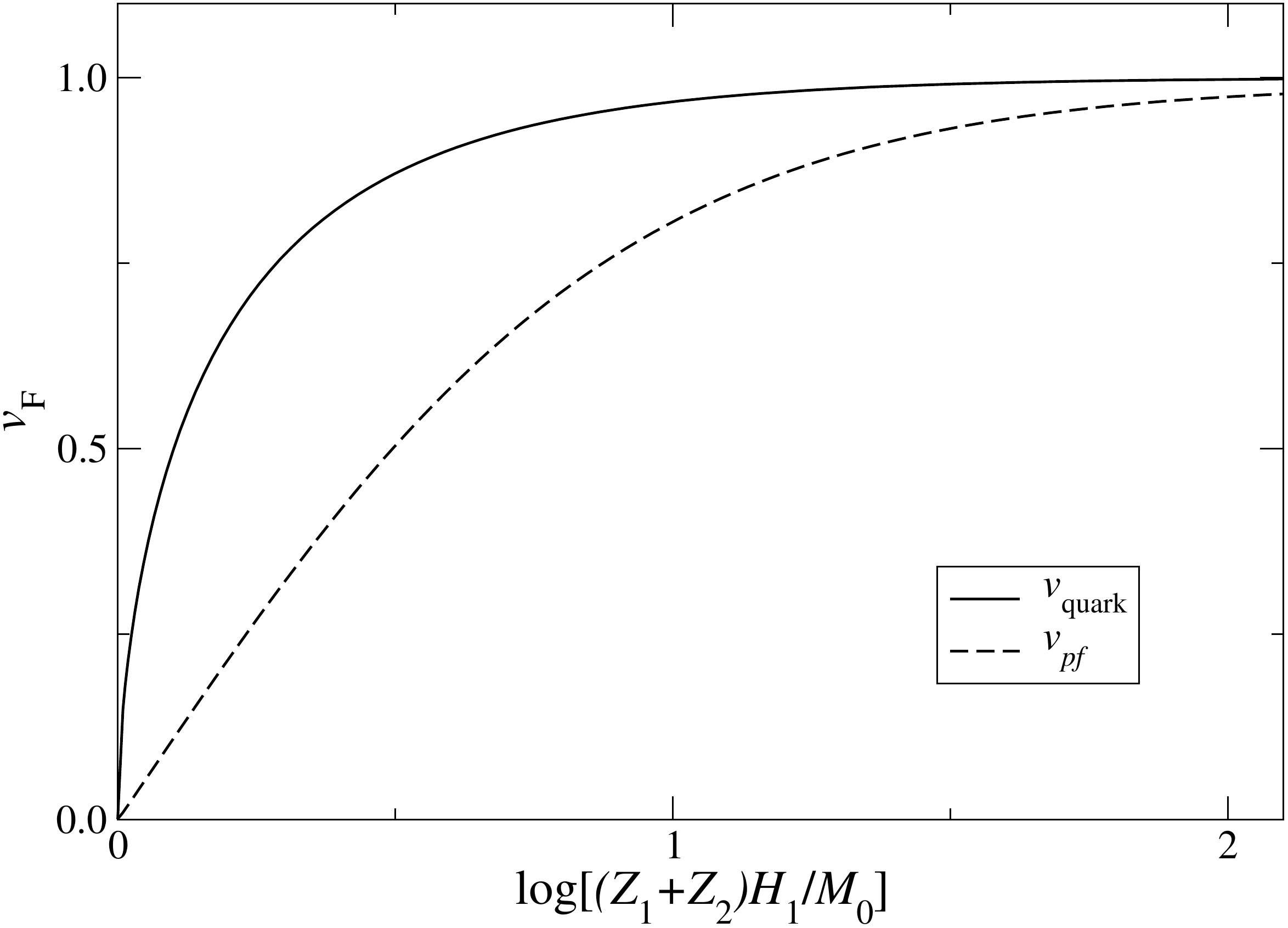}

    \caption{Fermi velocities of the quarks and first level $SU(3)$
      parafermion modes as a function of the field $Z_1H_1>M_0$ for
      $p_0=2+1/3$, $H_2\equiv 0$ at zero temperature.  For large fields,
      $H_1>H_{1,\delta}$, both Fermi velocities approach $1$ leading to the
      asymptotic result for the low temperature entropy (\ref{Entropy_CFT1}). 
    \label{fig:FermiV2}}
\end{figure}

\begin{figure}
  \includegraphics[width=0.6\textwidth]{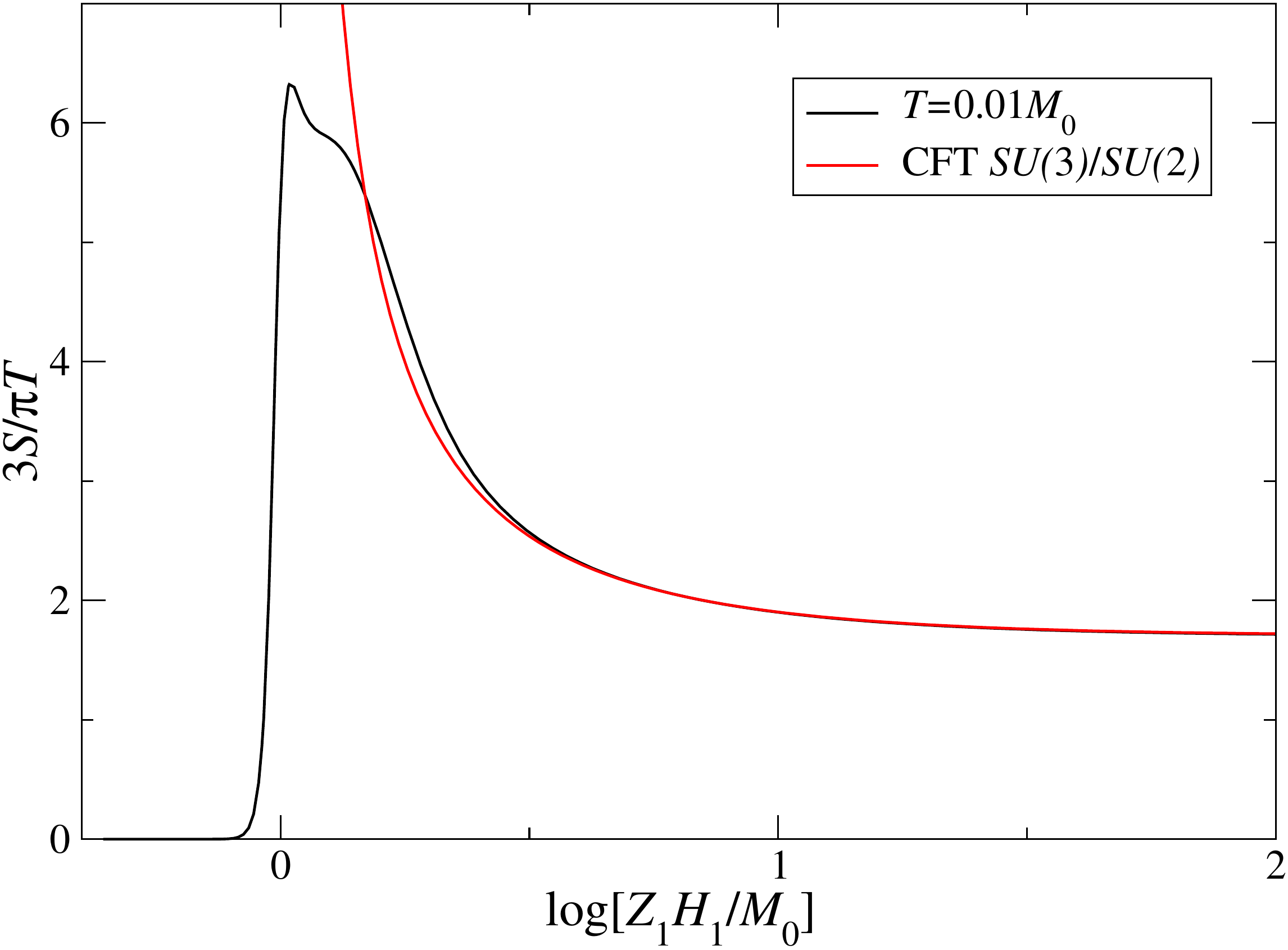}
  
  \caption{ Entropy obtained from numerical solution of the TBA equations
    (\ref{dressedeinteq}) for $p_0=2+1/3$ and $H_2\equiv 0$ as a function of
    the field $(Z_1H_1/M_0$ for $T=0.01M_0$. For fields large compared to the
    soliton mass, $Z_1H_1\gg M_0$, the entropy approaches the expected
    analytical value (\ref{Entropy_CFT1}) for a field theory with a free
    bosonic sector and a $Z_{SU(3)_{N_f=2}}/Z_{SU(2)_{N_f=2}}$ parafermion
    sector propagating with velocities $v^{(1)}_{\text{quark}}$ and
    $v^{(1)}_{pf}$, respectively (full red line). For magnetic fields
    $Z_1H_1<M_0$ and temperature $T\ll M_0$ the entropy is that of a dilute
    gas of non-interacting quasi-particles with degenerate internal degree of
    freedom due to the anyons.
    \label{fig:entropy2}}
\end{figure}

The resulting low temperature entropy is the sum of contributions from a
$U(1)$ boson (quark) and a $Z_{SU(3)_{N_f}}/Z_{SU(2)_{N_f}}$ parafermionic
coset (from the auxiliary modes)
\begin{equation}
\label{Ent_inter1} \mathcal{S} = \frac{\pi}{3} \left(
\frac{1}{v^{(1)}_\text{quark}} +
\frac{1}{v^{(1)}_{pf}}\left(\frac{6(N_f-1)}{N_f+3}-\frac{2(N_f-1)}{N_f+2}\right)
\right) T\,.
\end{equation} This behavior is consistent with the conformal embedding
(\ref{embedding1}).  Note that both Fermi velocities depend on the field $H_1$
and approach $1$ as $H_1\gtrsim H_{1,\delta}$, see Fig.~\ref{fig:FermiV2}.
Therefore, in this limit the entropy (\ref{Entropy_CFT1}) of the coset
$SU(3)_{N_f}/SU(2)_{N_f}$ is approached.  In Fig.~\ref{fig:entropy2} the
computed entropy (\ref{entropy_num}) of the model with $N_f=2$, $\nu=3$ is
shown for $T=0.01\,M_0$ as a function of the field $H_1$ together with the
$T\to0$ behaviour (\ref{Ent_inter1}) expected from conformal field theory.

\subsection{Condensate of quarks and antiquarks}
For fields $H_1\geq H_2 \gtrsim M_0/(Z_1+Z_2)$ the system forms a collective
state of solitons ($j=j_0$) on both levels, $m=1,2$.  Again, the low
temperature thermodynamics can be studied analytically for
$H_1 \ge H_2 \gg M_0/(Z_1+Z_2)$ .  Repeating the analysis of
Section~\ref{sectionB} we conclude from Eqs.~(\ref{dressedeinteq}),
(\ref{auxinteq}) that
\begin{equation}
  \begin{aligned}
    &f(\phi^{(m)}_{j_0}(\lambda_\delta))= \begin{cases}
      1 \quad m=1,\\
      1 \quad m=2
    \end{cases},
    \qquad f(\phi^{(m)}_{j_0}(\infty))=0,\\
    &f(\phi^{(m)}_{j_1}(\lambda_\delta))=0, \qquad\qquad\qquad\qquad\, f(\phi^{(m)}_{j_1}(\infty))=0,\\
    &f(\phi^{(m)}_{j_2}(\lambda_\delta))=
    \begin{cases}
      0 \qquad m=1,\\
      0 \qquad m=2
    \end{cases},\quad\,
    f(\phi^{(m)}_{j_2}(\infty))=\frac{\sin(\frac{\pi}{N_f+3})\sin(\frac{2\pi}{N_f+3})}{\sin(\frac{j_2\pi}{N_f+3})\sin(\frac{(j_2+1)\pi}{N_f+3})},
  \end{aligned}
\end{equation}
and therefore
\begin{equation}
  \rho^{h(m)}_{j_0}(\lambda)=\rho^{(m)}_{j_1}(\lambda)=\rho^{(m)}_{j_2}=0,
  \quad \mathrm{for~} |\lambda|<\lambda_\delta\,,
\end{equation}
giving $\mathcal{S}^{(m)}_j(\lambda_\delta)=0$ for all $j,m$. Using the
relation (\ref{RogerF}) the low-temperature behavior of the entropy is found
to be
\begin{equation}
  \label{Entropy_CFT2}
  \mathcal{S}=\frac{\pi}{3}\left(2+\frac{6(N_f-1)}{N_f+3} \right)T=\frac{\pi}{3}\frac{8N_f}{N_f+3}T
\end{equation}
in the phase with finite soliton density on both levels.  The low energy
excitations near the Fermi points $\epsilon^{(m)}_{j_0}(\pm\Lambda_m)=0$ of
the soliton dispersion propagate with velocities
$v^{(m)}_{\text{quark/antiquark}} = \left. (\partial_\lambda
  \epsilon^{(1/2)}_{j_0}) / (2\pi\rho^{(1/2)}_{j_0})\right|_{\Lambda_{1/2}}
\to 1$ for fields $H_m>H_{m,\delta}$ such that
$\Lambda_m(H_{m,\delta})>\lambda_\delta$.  From this we conclude that the
conformal field theory (CFT) describing the collective low energy modes is the
$SU(3)$ WZNW model at level $N_f$ or, by conformal embedding \cite{Gepner87},
a product of two free $U(1)$ bosons (contributing $c=2$ to the central charge)
and the $SU(3)$ parafermionic coset $SU(3)/U(1)^2$ with central charge
\begin{equation}
    \label{cSU3para}
	c=\frac{6(N_f-1)}{N_f+3}\,.
\end{equation}

\begin{figure}
    \includegraphics[width=0.6\textwidth]{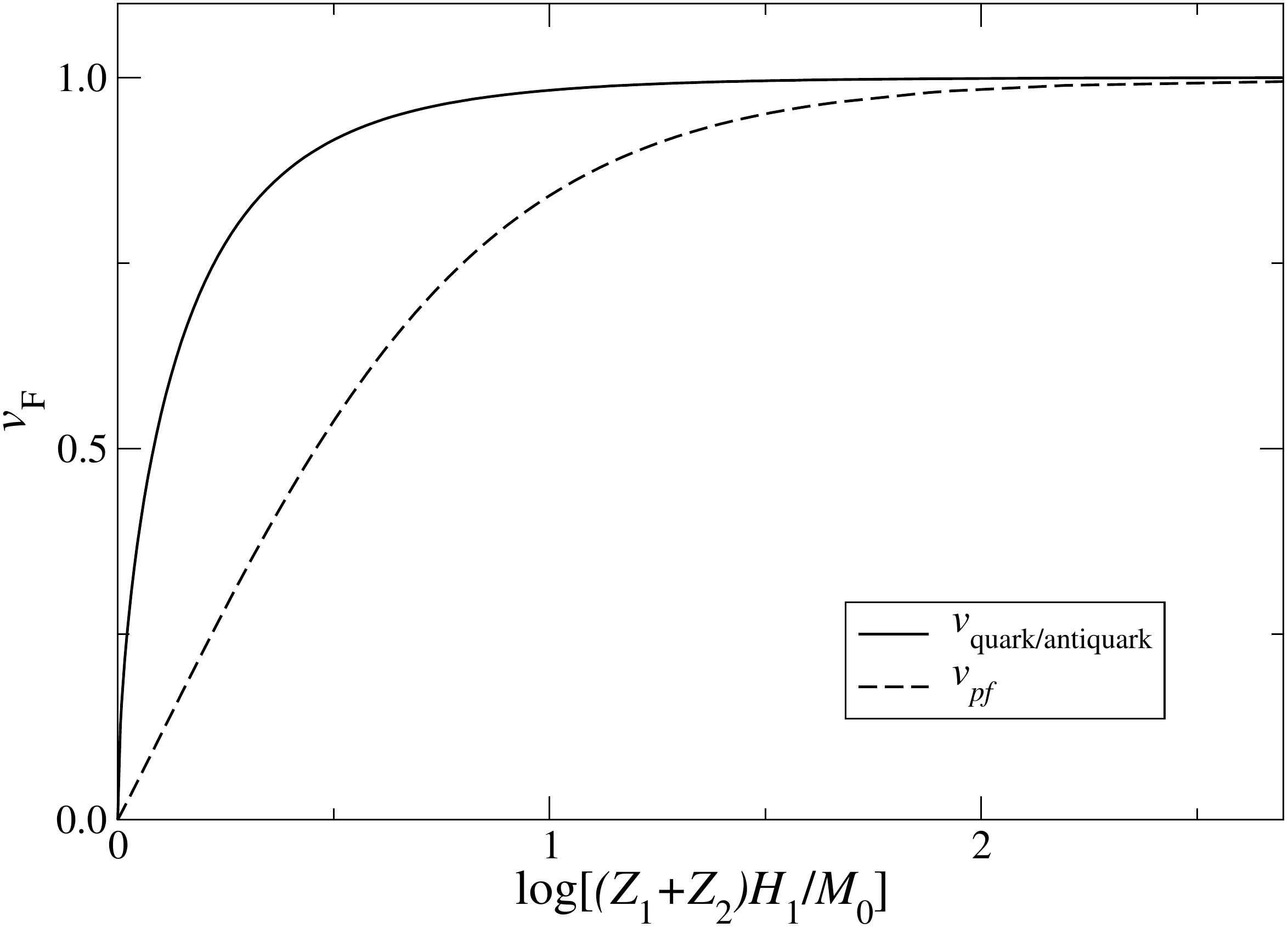}
  
  \caption{
        Fermi velocities of solitons and $SU(3)$ parafermion modes as a
        function of the magnetic field $H_1\equiv H_2 >M_0(Z_1+Z_2)$ for
        $p_0=2+1/3$ at zero temperature.  For large fields,
        $H_m>H_{m,\delta}$, both Fermi velocities approach $1$ leading to the
        asymptotic result for the low temperature entropy
        (\ref{Entropy_CFT2}). 
    \label{fig:FermiV}}
\end{figure}
\begin{figure}
    \includegraphics[width=0.6\textwidth]{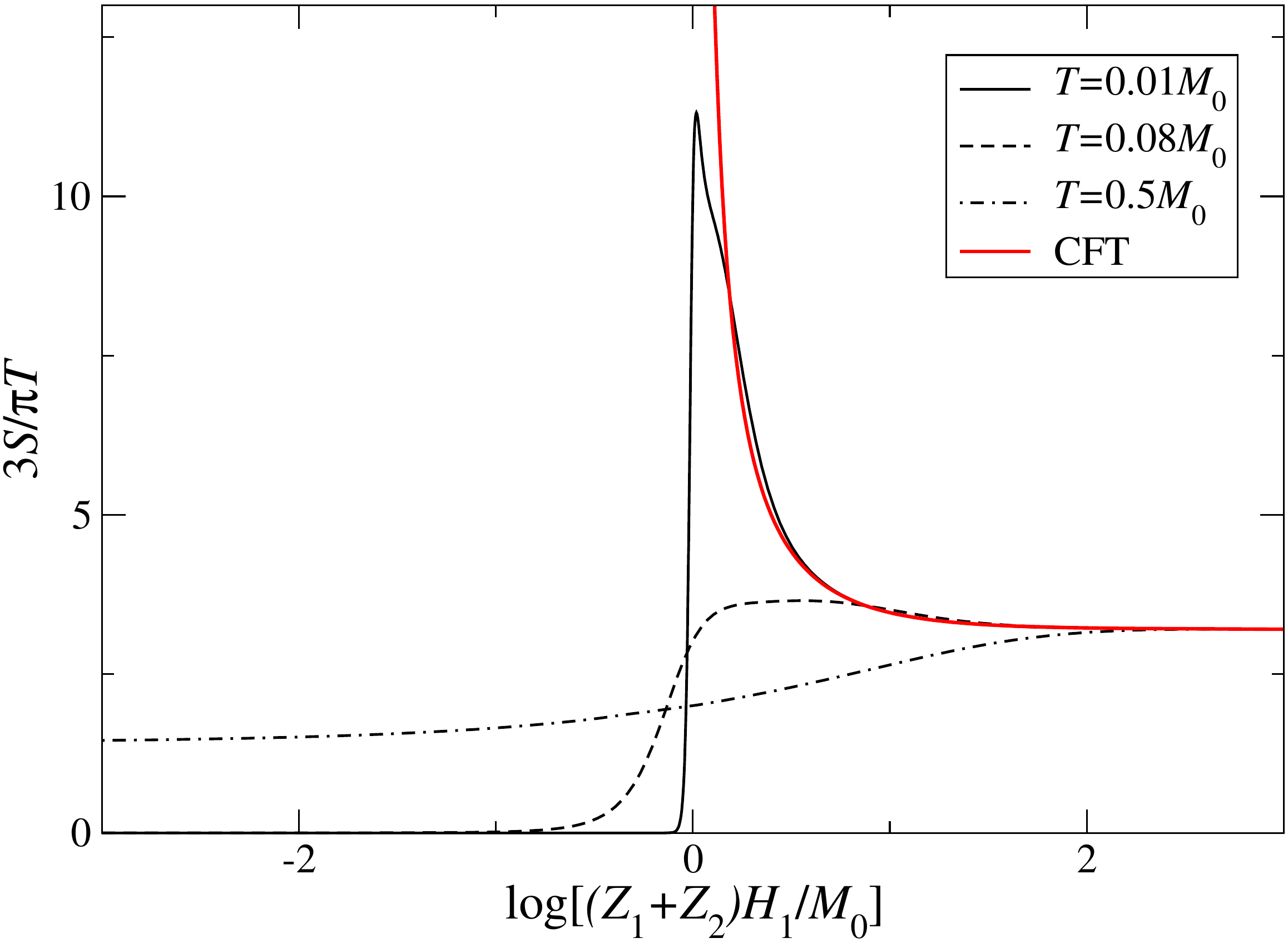}
  
  \caption{
        Same as Fig.~\ref{fig:entropy2} but for $H_1\equiv H_2$.  For fields
        large compared to the soliton mass, $(Z_1+Z_2)H_1\gg M_0$, the entropy
        approaches the expected analytical value (\ref{Entropy_CFT2}) for a
        field theory with two free bosonic sectors and a $SU(3)$ parafermion
        sector propagating with velocities
        $v_{\text{quark}}=v_{\text{antiquark}}$ and $v_{pf}$, respectively
        (full red line).  For magnetic fields $zH<M_0$ and temperature $T\ll
        M_0$ the entropy is that of a dilute gas of non-interacting
        quasi-particles with degenerate internal degree of freedom due to the
        $SU(3)_{N_f=2}$ anyons bound to the solitons. 
        \label{fig:Entropy1}}
\end{figure}

For the transition between this regime and the phases where the antiquarks are
gapped, see Section~\ref{sectionB}, we have to resort to an numerical analysis
of the TBA equations (\ref{dressedeinteq}) again:
in the case of equal fields, $H_1\equiv H_2$, where the corresponding modes on
the two levels are degenerate we find that the solitons $\epsilon_{j_0}^{(m)}$
propagate with Fermi velocity $v_{\text{quark}}$ while the auxiliary modes
$\epsilon_{j_2}^{(m)}$ propagate with velocity $v_{pf}$ (independent of
$j_2=1,\ldots,N_f-1$), see Fig.~\ref{fig:FermiV}.
%
%
As a consequence the contribution of the bosonic (quark/antiquark) and
parafermionic degrees of freedom to the low temperature entropy separate into
\begin{equation}
\label{Ent_inter}
    \mathcal{S} = \frac{\pi}{3} \left( \frac{2}{v_\text{quark}} + \frac{1}{v_{pf}}\frac{6(N_f-1)}{N_f+3}
    \right) T\,.
\end{equation}
In Fig.~\ref{fig:Entropy1} the computed entropy (\ref{entropy_num}) for the
model $SU(3)_{N_f=2}$ model with $\nu=3$ is shown for various temperatures as
a function of the fields $H_1\equiv H_2$ together with the $T\to0$ behaviour
(\ref{Entropy_CFT2}) expected from conformal field theory.
%

\begin{figure}
    \includegraphics[width=0.6\textwidth]{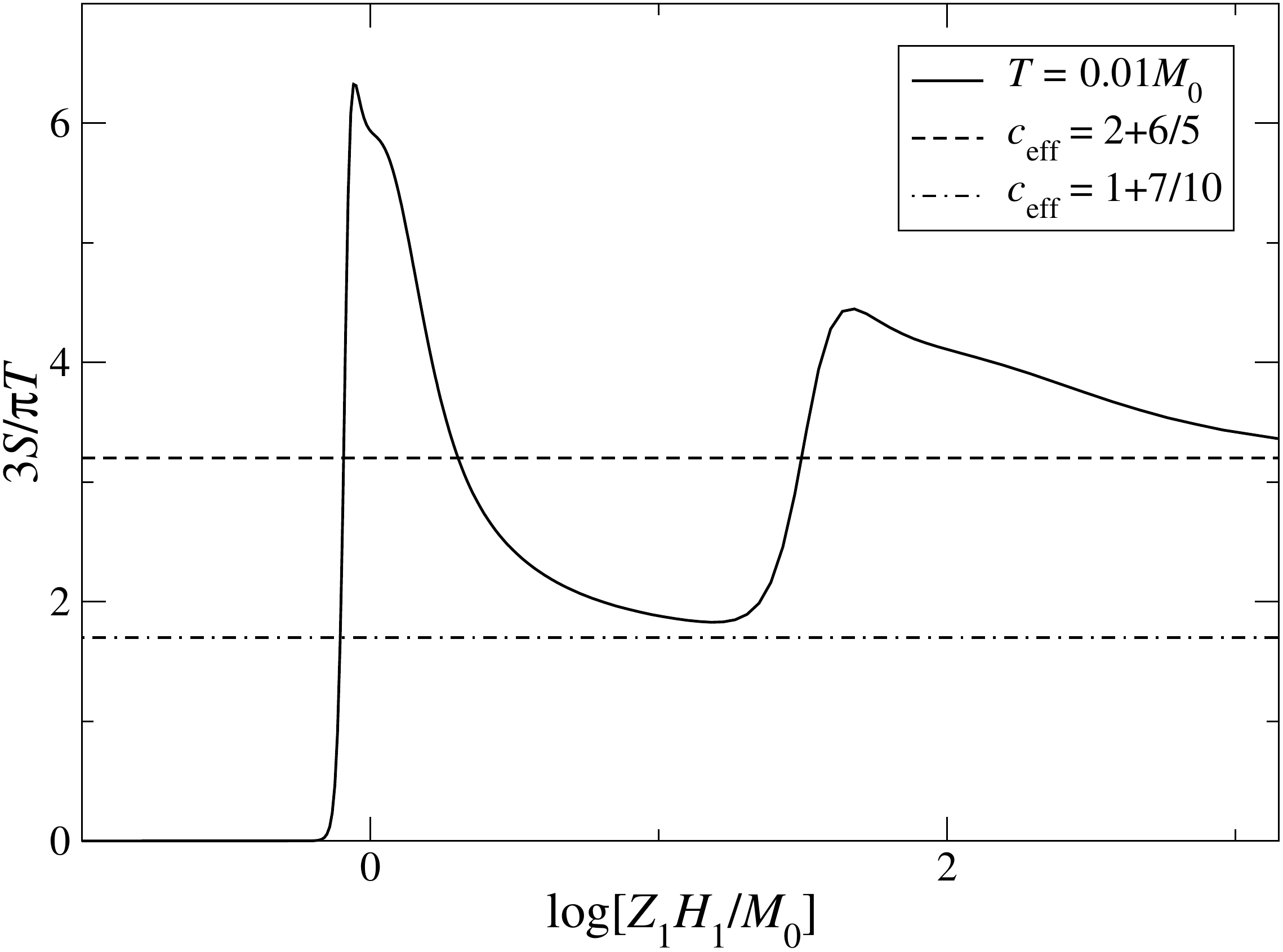}
  
  \caption{
        Same as Figure~\ref{fig:entropy2} but for fields along the line
        $\log(Z_2H_2/M_0)\equiv 0.4\log(Z_1H_1/M_0)-2.58$.  
        Transitions are observed at fields where the quarks and antiquarks
        condense, i.e.\ from the low density phase of non-interacting anyons
        into the collective state described by the
        $SU(3)_{N_f=2}/SU(2)_{N_f=2}$ coset CFT at and later into a phase
        whose low energy description is in terms of the  $SU(3)_{N_f=2}$ WZNW
        model.  The dashed-dotted lines indicate the corresponding central
        charges. 
        \label{fig:Entropy2}}
\end{figure}

Finally, in the regime $H_1\geq H_2 \gtrsim M_0/(Z_1+Z_2)$ the remaining
degeneracy between the two levels is lifted.  Quarks, antiquarks, and
auxiliary modes are propagating with (generically) different Fermi velocities
$v_{\text{quark}}$, $v_{\text{antiquark}}$, $v^{(1)}_{pf}$, and
$v^{(2)}_{pf}$.  The resulting low-temperature entropy behavior is found to be
\begin{equation}
  \begin{aligned}
    \mathcal{S}&=\frac{\pi}{3}\left(\frac{1}{v_{\text{quark}}}+\frac{1}{v_{\text{antiquark}}}+\frac{c_{1}}{v^{(1)}_{pf}}+\frac{c_{2}}{v^{(2)}_{pf}} \right)T,\\
    c_1&=\frac{6(N_f-1)}{N_f+3}-\frac{2(N_f-1)}{N_f+2}\,,\quad
    c_2=\frac{2(N_f-1)}{N_f+2},
  \end{aligned}
\end{equation}
consistent with the conformal embedding \cite{CFKM00}
\begin{equation}
  SU(3)_{N_f}=U(1)^2+Z_{SU(2)_{N_f}}+\frac{Z_{SU(3)_{N_f}}}{Z_{SU(2)_{N_f}}}.
\end{equation}
Fig.~\ref{fig:Entropy2} shows the transition between the different regimes described above.
%

\section{Summary and Conclusion}
\begin{figure}
  \includegraphics[width=0.85\textwidth]{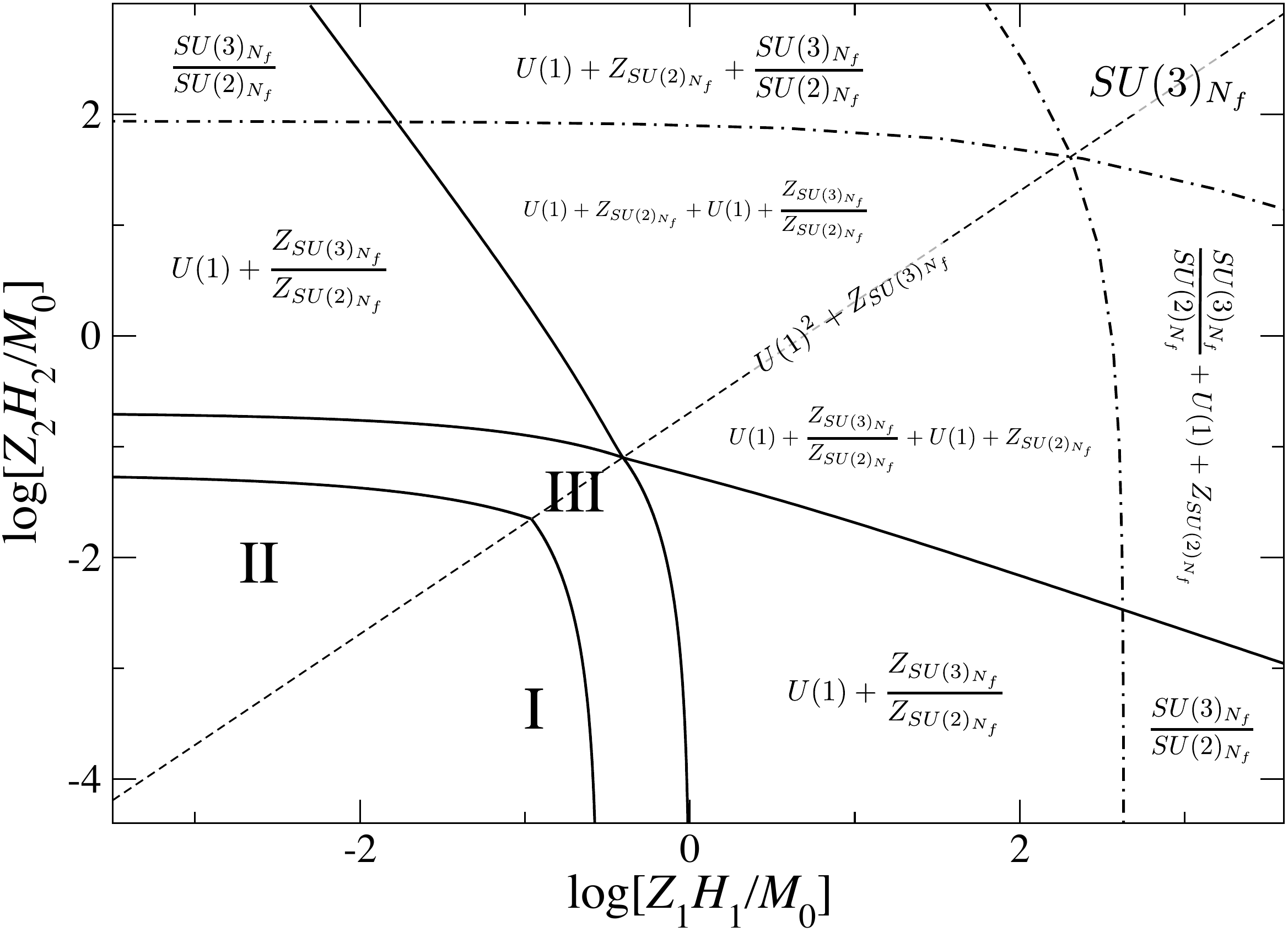}
  \caption{
    Contribution of the $SU(3)_{N_f}$ anyons to the low temperature properties
    of the model (\ref{NfModel}): using the criteria described 
       in the main text the parameter regions are identified using analytical
       arguments for $T\to0$ (the actual location of the boundaries is based
       on numerical data for $p_0=2+1/3$ and $T=0.05/M_0$). For small fields
       (regions I,II) a gas of non-interacting quasi-particles with the anyon
       as an internal zero energy degree of freedom bound to them is realized.
       Here the dashed line ($H_1\equiv H_2$ or $\epsilon^{(1)}_j\equiv
       \epsilon^{(2)}_j$) indicates the crossover between regions where the
       quarks (region I), or antiquarks (region II) dominate the free
       energy. In region III the presence of thermally activated solitons with
       a small but finite density lifts the degeneracy of the zero modes. The
       collective phases formed by condensed solitons are labelled by the
       corresponding CFTs providing the effective description of the low
       energy excitations. 
       \label{fig:phasediag}}
\end{figure}

We have studied the low temperature behaviour realized in a perturbed
$SU(3)_{N_f}$ WZNW model based on the Bethe ansatz solution of the model
describing the color sector of interacting fermions carrying color and flavor
degrees of freedom.  For small magnetic fields coupled to the $SU(3)$ charges
the elementary excitations form quark and antiquark multiplets with finite
mass carrying an internal non-Abelian $SU(3)_{N_f}$ anyonic degree of freedom.
Varying the magnetic field quarks and antiquarks condense.  As a result the
non-Abelian degrees of freedom bound to the the soliton excitations begin to
overlap.  Their resulting interaction lifts the degeneracy of these modes
resulting in the formation of various phases with propagating collective
excitations.  The effective theories describing these phases are products of
Gaussian fields for the soliton degrees of freedom and parafermionic cosets.
The latter describe the collective behaviour of interacting $SU(3)_{N_f}$
anyons related to the symmetries of the model.  Our findings are summarized in
Figure~\ref{fig:phasediag}.

\begin{acknowledgments}
  Funding for this work has been provided by the \emph{School for Contacts in
    Nanosystems}. HF gratefully acknowledges support from the by the Erwin
  Schr\"odinger Institute where part of this work has been done during the
  \emph{Quantum Paths} programme.  Additional support has been provided by the
  research unit \emph{Correlations in Integrable Quantum Many-Body Systems}
  (FOR2316).
\end{acknowledgments}

\appendix

\section{Thermodynamic Bethe ansatz}
\label{app:onTBA}
The fields $H_1,H_2$ appearing in (\ref{energy}) are defined by
$H_1\equiv\vec{\alpha}_1\cdot\vec{h}$, $H_2\equiv\vec{\alpha}_2\cdot\vec{h}$,
where the fields $\vec{h}=(h_1,h_2)$ couple to the generators of the Cartan
subalgebra of $SU(3)$ and $\vec{\alpha}_1,\,\vec{\alpha}_2$ denote the simple
roots of $SU(3)$.

In order to obtain the integral equations (\ref{densities2}) we consider a
root configuration consisting of $\nu^{(m)}_j$ strings of type $(n_j,v_{n_j})$
on the $m-$th level and using (\ref{string}) the Bethe equations
(\ref{betheeq}) can be rewritten in terms of the real string-centers
$\lambda^{(m,j)}_\alpha \equiv \lambda^{(m)n_j}_\alpha$.  In their logarithmic
form they read
\begin{equation}
  \label{centereq}
  \begin{aligned} \sum_{\tau=\pm
1}\frac{\mathcal{N}}{2}t_{k,N_f}(\lambda^{(1,k)}_\alpha+\tau/g)&=2\pi
I^{(1,k)}_\alpha+\sum_j\sum_{\beta=1}^{\nu^{(1)}_j}\theta^{(1)}_{kj}(\lambda^{(1,k)}_\alpha-\lambda^{(1,j)}_\beta)-\sum_j\sum_{\beta=1}^{\nu^{(2)}_j}\theta^{(2)}_{kj}(\lambda^{(1,k)}_\alpha-\lambda^{(2,j)}_\beta),\\
0 &= 2\pi I^{(2,k)}_\alpha
+\sum_j\sum_{\beta=1}^{\nu^{(2)}_j}\theta^{(1)}_{kj}(\lambda^{(2,k)}_\alpha-\lambda^{(2,j)}_\beta)-\sum_j\sum_{\beta=1}^{\nu^{(1)}_j}\theta^{(2)}_{kj}(\lambda^{(2,k)}_\alpha-\lambda^{(1,j)}_\beta),
  \end{aligned}
\end{equation}
where $I^{(m,k)}_\alpha$ are integers (or half-integers) and we have
introduced the functions
\begin{equation}
  \label{kernelrel1}
  \begin{aligned}
    t_{k,N_f}(\lambda)&=\sum_{l=1}^{\min(n_k,N_f)}f(\lambda,|n_k-N_f|+2l-1,v_kv_{N_f}),\\
    \theta^{(1)}_{kj}(\lambda)&=f(\lambda,|n_k-n_j|,v_kv_j)+f(\lambda,n_k+n_j,v_kv_j)+2\sum_{\ell=1}^{\min(n_k,n_j)-1}f(\lambda,|n_k-n_j|+2\ell,v_kv_j),\\
    \theta^{(2)}_{kj}(\lambda)&=\sum_{l=1}^{\min(n_k,n_j)}f(\lambda,|n_k-n_j|+2l-1,v_kv_j)
  \end{aligned}
\end{equation}
with
\begin{equation*}
  f(\lambda,n,v)= \begin{cases}
    2\arctan\left(\tan((\frac{1+v}{4}-\frac{n}{2p_0})\pi)\tanh(\frac{\pi\lambda}{2p_0})\right)
    \quad \text{if }\frac{n}{p_0}\neq \text{integer} \\ 0 \hspace*{6.8cm} \text{if
    }\frac{n}{p_0}= \text{integer}
  \end{cases}\,.
\end{equation*}
In the thermodynamic limit, $N_m,\mathcal{N}\rightarrow \infty$ with
$N_m/\mathcal{N}$ fixed, the centers $\lambda^{(m,k)}_\alpha$ are distributed
continuously with densities $\rho^{(m)}_k(\lambda)$ and hole densities
$\rho^{h(m)}_k(\lambda)$. Following \cite{YaYa66b} the densities are defined
through the following integral equations
\begin{equation}
  \label{densities1}
  \begin{aligned}
    \tilde{\rho}^{(1)}_{0,k}(\lambda) &= (-1)^{r(k)}\rho^{h(1)}_k(\lambda)+\sum_jA^{(1)}_{kj}\ast \rho^{(1)}_j(\lambda)-\sum_jA^{(2)}_{kj}\ast\rho^{(2)}_j(\lambda),\\
    0 &= (-1)^{r(k)}\rho^{h(2)}_k(\lambda)+\sum_jA^{(1)}_{kj}\ast \rho^{(2)}_j(\lambda)-\sum_jA^{(2)}_{kj}\ast\rho^{(1)}_j(\lambda),
  \end{aligned}
\end{equation}
where $a\ast b$ denotes a convolution and $r(j)$ is defined in Appendix A of
\cite{BoFr18}. The bare densities
$\tilde{\rho}^{(1)}_{0,j}(\lambda)$ and the kernels $A^{(m)}_{jk}(\lambda)$ of
the integral equations are defined by
\begin{equation}
  \label{kernelrel2}
  \begin{aligned}
    \tilde{\rho}^{(1)}_{0,j}(\lambda)&=\frac{1}{2}\left(a_{j,N_f}(\lambda+1/g)+a_{j,N_f}(\lambda-1/g)\right), \quad
    a_{j,N_f}(\lambda)=\frac{1}{2\pi}\frac{d}{d\lambda}t_{j,N_f}(\lambda)\\
    A^{(m)}_{kj}(\lambda)&=\frac{1}{2\pi}\frac{d}{d\lambda}\theta^{(m)}_{kj}(\lambda)+(-1)^{r(k)}\delta_{m1}\delta_{jk}\delta(\lambda)\,.
  \end{aligned}
\end{equation}
We rewrite the energy density $\mathcal{E}=E/\mathcal{N}$ using (\ref{energy})
and the solutions $\rho^{(m)}_k$ of (\ref{densities1}) as
\begin{equation}
  \label{energy2}
  \begin{aligned}
    \mathcal{E}&=\frac{1}{\mathcal{N}}\sum_{j}\sum_{\alpha=1}^{\nu^{(1)}_j}\left(\sum_{\tau=\pm
        1}\frac{\tau}{2}t_{j,N_f}(\lambda^{(j,1)}_{\alpha}+\tau/g)+n_jH_1\right)+\frac{1}{\mathcal{N}}\sum_j\sum_{\alpha=1}^{\nu^{(2)}_j}n_jH_2-\frac{2}{3}H_1-\frac{1}{3}H_2\\
    &\stackrel{\mathcal{N}\rightarrow \infty}{=}\sum^2_{m=1}\sum_{j\geq
      1}\int_{-\infty}^{+\infty}\text{d}\lambda\,
    \tilde{\epsilon}^{(m)}_{0,j}(\lambda)\rho^{(m)}_{j}(\lambda)-\frac{2}{3}H_1-\frac{1}{3}H_2,
  \end{aligned}
\end{equation}
where we introduced the bare energies
\begin{equation}
  \tilde{\epsilon}^{(1)}_{0,j}(\lambda)=\sum_{\tau=\pm 1}\frac{\tau}{2}t_{j,N_f}(\lambda+\tau/g)+n_jH_1, \quad
  \tilde{\epsilon}^{(2)}_{0,j}(\lambda)=n_jH_2.
\end{equation}
It turns out that the energy (\ref{energy2}) is minimized by a configuration,
where only the strings of length $N_f$ have a finite density. After inverting
the kernels $A^{(1)}_{j_0j_0}$ on both levels in equation (\ref{densities1})
and inserting the resulting expression for $\rho^{(m)}_{j_0}(\lambda)$ into
the other equations for $k\neq j_0$ we end up with the integral equations
(\ref{densities2}), where we redefined the densities
$\rho^{h(m)}_{j_0}\leftrightarrow \rho^{(m)}_{j_0}$ and introduced the kernels
$B^{(1,1)}_{jk}=B^{(2,2)}_{jk}\equiv B^{(1)}_{jk}$,
$B^{(1,2)}_{jk}=B^{(2,1)}_{jk}\equiv B^{(2)}_{jk}$, whose Fourier transformed
kernels $B^{(1)}_{jk}(\omega)$, $B^{(2)}_{jk}(\omega)$ are given by
\begin{equation}
  \begin{aligned}
    B^{(m)}_{j_0j_0}&=\frac{A^{(m)}_{j_0j_0}}{(A^{(1)}_{j_0j_0})^2-(A^{(2)}_{j_0j_0})^2},\\
    B^{(m)}_{kj_0}&=(-1)^{r(k)}\frac{A^{(2)}_{kj_0}A^{((m \text{ mod 2})+1)}_{j_0j_0}-A^{(1)}_{kj_0}A^{(m)}_{j_0j_0}}{(A^{(1)}_{j_0j_0})^2-(A^{(2)}_{j_0j_0})^2}, \qquad k\neq j_0,\\
    B^{(m)}_{j_0k}&=-(-1)^{r(k)}B^{(m)}_{kj_0}, \qquad k\neq j_0,\\
    B^{(m)}_{kj}&=(-1)^{r(k)}\left(A^{(2)}_{kj_0}B^{((m \text{ mod }2)+1)}_{j_0j}-A^{(1)}_{kj_0}B^{(m)}_{j_0j}+(-1)^{m+1}A^{(m)}_{kj}\right), \qquad k,j \neq j_0.
  \end{aligned}
\end{equation}
The inverse of the Fourier transformed kernel $B^{(1)}_{j_2k_2}(\omega)$ is
given by $C_{j_2k_2}=\delta_{j_2k_2}-s(\delta_{j_2k_2+1}+\delta_{j_2k_2-1})$
\cite{Tsve87}. Hence, the Fourier transformed kernels $C^{(m)}_j(\omega)$ of
equation (\ref{auxinteq}) are defined by
\begin{equation}
  \sum_{j_2}C_{k j_2}B^{(1)}_{jj_2}=\delta_{k,N_f-1}C^{(m)}_j.
\end{equation}

\newpage

%

\end{document}